\documentclass[aps,prd,preprint,groupedaddress]{revtex4-1}
\usepackage{amsfonts,amsmath,amssymb}
\usepackage{color}
\usepackage{hyperref}

\def\exd{\mathrm{d}}
\def\exD{\mathrm{D}}

\def\hphi{{\phi}}
\def\hphid{{{{\phi}}^\dag}}

\begin{document}
\title{Emergent gravitational dynamics in relativistic Bose--Einstein condensate}
\author{Alessio Belenchia}
\email[]{alessio.belenchia@sissa.it}
\author{Stefano Liberati}
\email[]{stefano.liberati@sissa.it}
\author{Arif Mohd}
\email[]{arif.mohd@sissa.it}
\affiliation{SISSA - International School for Advanced Studies, \\
Via Bonomea 265, 34136 Trieste, Italy \\
and \\
INFN, Sezione di Trieste, Trieste, Italy.}
\date{\today}
\begin{abstract}
Analogue models of gravity have played a pivotal role in the past years by providing a test bench for many open issues in quantum field theory in curved spacetime such as the robustness of Hawking radiation and cosmological particle production. More recently, the same models have offered a valuable framework within which current ideas about the emergence of spacetime and its dynamics could be discussed via convenient toy models. In this context, we study here an analogue gravity system based on a relativistic Bose--Einstein condensate. We show that in a suitable limit this system provides not only an example of an emergent spacetime (with a massive and a massless relativistic fields propagating on it) but also that such spacetime is governed by an equation with geometric meaning that takes the familiar form of Nordstr{\"o}m theory of gravitation. In this equation the gravitational field is sourced by the expectation value of the trace of the effective stress energy tensor of 
 the quasiparticles while the Newton and cosmological constants are functions of the fundamental scales of the microscopic system. This is the first example of analogue gravity in which a Lorentz invariant, geometric theory of semiclassical gravity emerges from an underlying quantum theory of matter in flat spacetime.\\
\end{abstract}
\maketitle
\tableofcontents
\section{Introduction}
In recent years there has been considerable interest in emergent gravity scenarios which envision that general relativity and spacetime itself could be a sort of thermodynamic limit of a more fundamental theory based on  fundamental entities different from spacetime quanta. The links between thermodynamics and gravitation --  black-hole thermodynamics \cite{Bardeen:1973gs,Hawking:1974sw,Hawking:1974rv}; derivation of the Einstein equation from the thermodynamics of local causal horizons \cite{Jacobson:1995ab} (see however~\cite{Chirco:2014saa} for different point of view); membrane description of event horizon \cite{Thorne:1986iy, Damour:1978cg} -- are the underlying motivations behind the emergent gravity program. \par
While this paradigm might seem at odd with some approaches to quantum gravity it is not necessarily incompatible with them. In particular,  emergence akin to that in the condensed-matter systems have been increasingly studied within the quantum gravity community in order to understand the emergence of spacetime and gravity from different fundamental ontologies~\cite{Sindoni:2011ej,Gielen:2013naa,Oriti:2007qd,Hu:2005ub,Dreyer:2007ws,Dreyer:2006pp} (see also~\cite{Barcelo:2014yna} for a recent extension of these ideas to electromagnetism). In this sense, emergent gravity settings might end up being more a completion of quantum gravity scenarios rather than a drastic alternative. \par

Analogue models of gravity are provided by several condensed-matter/optical systems in which the excitations propagate in a relativistic fashion on an emergent pseudo-Riemannian geometry induced by the medium. Since the seminal work of Unruh \cite{Unruh:1980cg} analogue models of gravity have set a fruitful arena in which issues related to semi-classical gravity can be studied in concrete toy models (see for e.g.~\cite{Liberati:2009uq, Barcelo:2005fc} and references therein). While the main focus in this area has been to experimentally simulate phenomena expected within quantum field theory on curved spacetime, e.g., analogue Hawking radiation \cite{Unruh:1980cg,Jacobson:1993hn,Barcelo:2003wu} and cosmological particle production \cite{Weinfurtner:2008if,Weinfurtner:2008ns}, it has also been shown that the emergence of a Lorentz signature metric is a characteristic of a large class of systems~\cite{Barcelo:2001tb} and can also be obtained starting from Euclidean field theories~\cite{Girelli:2008qp}. \par
Among the various analogue systems, a preeminent role has been  played by Bose--Einstein condensates (BEC) because these are macroscopic quantum systems whose phonons/quasi-particle excitations can be meaningfully treated quantum mechanically and hence used to fully simulate the above mentioned quantum phenomena~\cite{Garay:1999sk, Garay:2000jj, Barcelo:2003wu}.

 Most of the research on analogue gravity so far has dealt with the questions related to the emergence of spacetime and  quantum field theory on it. The analogue of  gravitational dynamics however is generally missing, i.e.~the spacetime that emerges has a dynamics which cannot be cast in the form of background independent geometric equations. Nonetheless, there are have been in recent times attempts of reproducing the emergence of some gravitational dynamics  within analogue gravity systems (see e.g.~\cite{Girelli:2008gc,Volovik:2013fca,Volovik:2009av,Jannes:2011em}).  

In particular, in a recent development, one of the present authors and collaborators succeeded in finding the analogue of the Poisson equation for the gravitational potential associated to the background geometry experienced by the quasi-particles propagating on a non-relativistic BEC~\cite{Sindoni:2009fc}. Noticeably this equation is sourced, as in Newtonian gravity, by the density of the quasi-particles (the analogue of the matter in this system) while a cosmological constant is also present due to the back-reaction of the atoms which are not part of the condensate (the so called depletion factor)~\cite{Sindoni:2009fc,Finazzi:2011zw}. While the appearance of an analogue gravitational dynamics in a BEC system is remarkable, it is not a surprise that this analogue system is able to produce only  Newtonian  gravity since the  model itself is based on the non-relativistic BEC to start with. Nonetheless a derivation of relativistic gravitational dynamics in analogue models has been missing so far. \par

Remarkably, BEC can also be described within a completely relativistic framework and indeed a relativistic BEC (rBEC) --- a Bose--Einstein condensate of a system of relativistic particles --- was first studied as an analogue model in~\cite{Fagnocchi:2010sn} where it was shown that the low-energy massless quasi-particles propagate as massless minimally coupled scalar field on a curved spacetime. The quasi-particles thus feel a curved effective metric called the acoustic metric. The dynamics of the acoustic metric itself, however, was not discussed in that work. It is natural to expect that a rBEC might provide a suitable model for the relativistic dynamics of an emergent spacetime. This is the subject of this paper. \par
The plan of the paper is as follows. We begin in sec.~\ref{sec:relBEC} with a review of Bose--Einstein condensation in a complex scalar field theory. In sec.~\ref{sec:analogue} we will study the dynamics of the condensate and will show how the perturbations experience the condensate as a curved spacetime geometry. In sec.~\ref{sec:relStBEC} we will make contact with previous work on relativistic BEC ~\cite{Fagnocchi:2010sn} stressing also the differences with respect to our work. This section can be skipped by readers not familiar with ref.~\cite{Fagnocchi:2010sn} as it is not strictly needed for the overall understanding of our results. In sec.~\ref{sec:emergence} we will show how the dynamics in this model can be interpreted as the emergence of a Lorentz invariant theory of gravity -- the Nordstr{\"o}m gravity. We conclude with a summary of results and outlook in sec.~\ref{sec:discuss}. \par
Our metric signature is $(-+++)$ and the conventions are those of Wald in ref.~\cite{Wald:1984rg}.
\section{Complex scalar field theory: relativistic BEC }
\label{sec:relBEC}
 Let us start by considering the general theory for a relativistic Bose--Einstein condensation. This is generically described by a complex scalar field endowed with an internal $U(1)$ symmetry which ends up to be spontaneously  broken below some critical temperature \cite{Kapusta:1981aa,Bernstein:1990kf,Haber:1981ts,Haber:1981fg}.
We shall closely follow~\cite{Kapusta:1981aa}, to which we refer the reader for a detailed treatment.

The Lagrangian is given by
\begin{equation}
\label{eq:Lagrangiana}
\mathcal{L}=-\eta^{\mu\nu}\partial_\mu \hphid \partial_\nu \hphi-m^2
\hphid\hphi-\lambda(\hphid \hphi)^2.
\end{equation}
The theory has a $U(1)$-invariance under phase rotation of the fields. The corresponding conserved current is given by
\begin{align}
j_\mu = i(\hphid \partial_\mu \hphi - \hphi \partial_\mu \hphid).
\end{align}
Space integral of the zeroth component of current gives the conserved charge,
\begin{align}
Q = i \int \exd^3 x (\hphid \partial_t \hphi - \hphi \partial_t \hphid).
\end{align}
To describe the theory at a finite temperature $T=1/\beta$ we Wick-rotate the time $\tau=-it$ and periodically identify the fields with a period $\tau=\beta$. Instead of using a complex field, it is convenient to use
the real and imaginary parts of $\phi$ as dynamical variables: $(\phi_1 + i \phi_2)/\sqrt{2}$. Defining the momentum conjugate to the fields as $\pi_i = \partial \phi_i / \partial t$ for $i=1,2$, the partition function at a finite value of charge is then given by
\begin{align}
\label{eq:partition fn}
\mathcal{Z} = \mathcal{N} \int \exD\pi_1 \exD\pi_2 \exD\phi_1 \exD\phi_2 \,\, \mathrm{exp} \left[\int_0^\beta \exd\tau \int \exd^3x \left( i \pi_1 \dot\phi_1 + i \pi_2 \dot\phi_2 - [\mathcal{H} - \mu \mathcal{Q}] \right)\right],
\end{align}
where $\mu$ is the chemical potential sourcing the charge density $\mathcal{Q}=\phi_2 \pi_1 - \phi_1 \pi_2$ in the system and $\mathcal{H}$ is the Hamiltonian density
\begin{align}
\mathcal{H} = \frac{1}{2}\left( \pi_1^2 + \pi_2^2 + (\vec{\nabla}\phi_1)^2 + (\vec{\nabla}\phi_2)^2 + m^2\left(\phi_1^2 + \phi_2^2\right)\right)+\frac{\lambda}{4}(\phi_1^2 + \phi_2^2)^2.
\end{align}
Total amount of  charge at equilibrium can be obtained from the partition function as
\begin{align}
\label{eq:net charge}
Q = \frac{1}{\beta} \frac{\partial}{\partial \mu} \text{ln} \, \mathcal{Z}.
\end{align}
In the laboratory, one prepares the system with some net amount of charge $Q$ and the value of $\mu$ is obtained by inverting eq.~\eqref{eq:net charge}. \par
The integral over momenta in eq.~\eqref{eq:partition fn} is a Gaussian integral. Hence, the momenta can be integrated away. This gives
\begin{align}
\mathcal{Z} = \mathcal{N}_\beta \int \exD\phi_1 \exD\phi_2 \,\, \mathrm{exp} \left[-\int_0^\beta \exd\tau \int \exd^3x \,\, \mathcal{L}_{\rm eff}\,\right],
\end{align}
where $\mathcal{N}_\beta$ is a $\beta$ dependent constant, and $\mathcal{L}_{\rm eff}$ is the effective Lagrangian of the theory given by
\begin{align}
\label{eq:eff L}
\mathcal{L}_{\rm eff} = \frac{1}{2} \left(\dot\phi_1^2 + \dot\phi_2^2 + (\vec{\nabla}\phi_1)^2 + (\vec{\nabla}\phi_2)^2\ \right) +
i \mu (\phi_2 \dot\phi_1 - \phi_1 \dot\phi_2) + V(\phi)
\end{align}
where $V(\phi)$ is the effective potential given by
\begin{align}
V(\phi)= \frac{1}{2}
(m^2-\mu^2) (\phi_1^2 + \phi_2^2) + \frac{\lambda}{4} (\phi_1^2 + \phi_2^2)^2
\end{align}
From the form of the effective potential it is clear that at a given $\beta$ if $\mu > m$ then the system is in the broken $U(1)$ phase and the condensate has formed. It can be shown \cite{Kapusta:1981aa} that this phase transition is second order and the critical temperature is given by
\begin{equation}\label{eq:Tcrit}
T_{c}=\frac{3}{\lambda}\left(\mu^{2}-m^{2}\right).
\end{equation}
Later we shall be interested in the massless limit for which the critical temperature is given by $T_c = 3 \mu^2/\lambda$. Thus, in the massless case, a non-zero chemical potential is necessary in order for the $U(1)$ symmetry to be broken and the condensate to be formed at a finite non-zero critical temperature. In the following we shall always consider the system to be at temperatures $T \ll T_c$, so that  thermal effects can be safely neglected.
\section{Relativistic BEC as an analogue gravity model}
\label{sec:analogue}
\subsection{Dynamics of the condensate: Gross--Pit{\ae}vskii equation}
\label{subsec:condensate}
The effective Lagrangian of eq.~\eqref{eq:eff L} can be rewritten in terms of the complex valued fields as
\begin{align}
\label{eq:Lagrangian}
\mathcal{L}_{\rm eff} = -\eta^{\mu \nu} \partial_\mu \phi^* \partial_\nu \phi - m^2 \phi^* \phi - \lambda (\phi^* \phi)^2 + \mu^2 \phi^* \phi + i \mu(\phi^* \partial_t \phi - \phi \partial_t \phi^*)
\end{align}
The equation of motion for $\phi$ is obtained by variation with respect to $\phi^*$ and we get,
\begin{align}
\left(-\Box + m^2 - \mu^2 - 2i\mu \frac{\partial}{\partial t}\right)\phi + 2 \lambda (\phi^*\phi) \phi = 0.
\end{align}
We can factor out explicitly the dependence on the chemical potential and write the field as
\begin{equation}
\phi=\varphi e^{i\mu t}.
\end{equation}
This gets rid of the $\mu$ dependent terms and we get
\begin{align}
\label{eq:master}
\left(\Box - m^2\right)\varphi - 2 \lambda |\varphi|^{2}\varphi = 0.
\end{align}
This was the starting equation in  ref.~\cite{Fagnocchi:2010sn} where the acoustic metric was first derived.
Let us now decompose $\varphi$ as $\varphi=\varphi_0(1+\psi)$, where $\varphi_0$ is the condensed part of the field $(\langle \varphi \rangle = \varphi_0)$, which we take to be real, and $\psi$ is the fractional fluctuation. {The reality of the condensate order parameter is the crucial assumption here. We will comment on this in the discussion section.} Note that $\psi$ is instead complex and $\langle \psi \rangle = 0 $. It can be written in terms of its real and imaginary parts $\psi = \psi_1 + i\psi_2$. Substituting this decomposition in eq.~\eqref{eq:master} and taking the expectation value we get the equation of motion for the condensate
\begin{align}
\label{eq:GP}
(\Box - m^2)\varphi_0 - 2 \lambda \varphi_0^3 - 2 \lambda \varphi_0^3
\, \left[\,3\, \langle\psi_1^2\rangle + \langle\psi_2^2\rangle\,\right]=0,
\end{align}
where we have assumed that the cross-correlation of the fluctuations vanish, i.e., $\langle \psi_1 \psi_2 \rangle = 0$. This is justified a posteriori by equations \eqref{eq:acousticfluctuations}, which show that $\psi_1$ and $\psi_2$ do not interact with each other at the order of approximation we are working. Eq.~\eqref{eq:GP} determines the dynamics of the condensate taking into account the backreaction of the fluctuations. It is the relativistic generalization of the Gross--Pit{\ae}vskii equation \cite{pitaevskii2003bose}.
\subsection{Dynamics of perturbations: acoustic metric }
\label{susbec:perturbations}
Having determined the dynamics of the condensate we now want to calculate
the equations of motion for the perturbations themselves.
To this end, we insert $\varphi=\varphi_0(1+\psi_1 + i\psi_2)$ in eq.~\eqref{eq:master} and expand it to linear order in $\psi$'s. Using the Gross--Pit{\ae}vskii equation to that order and separating the real and imaginary parts we get the equation of motion for $\psi_1$ and $\psi_2$,
\begin{subequations}
\label{eq:fluctuations}
\begin{align}
\Box \psi_1 + 2 \eta^{\mu \nu}\partial_\mu (\ln\varphi_0) \partial_\nu \psi_1 - 4 \lambda \varphi_0^2 \psi_1 = 0, \\
\Box \psi_2 + 2 \eta^{\mu \nu}\partial_\mu (\ln\varphi_0) \partial_\nu \psi_2 = 0.
\end{align}
\end{subequations}
We therefore see that $\psi_2$ is the massless mode, which is the Goldstone boson of the broken $U(1)$ symmetry, while $\psi_1$ is the massive mode with mass $2 \varphi_0\sqrt{\lambda}$. We now define a ``acoustic" metric, which is conformal to the background Minkowski,
\begin{align}
\label{eq:acousticmetric}
g_{\mu \nu} = \varphi_0^2 \, \eta_{\mu \nu}.
\end{align}
The relation between the d'Alembertian operators for $g_{\mu \nu}$ and $\eta_{\mu \nu}$ is given by,
\begin{align}
\Box_g = \frac{1}{\varphi_0^2}\,\Box + \frac{2}{\varphi_0^2}\, \eta^{\mu \nu}\,\partial_\mu (\ln\varphi_0)\, \partial_\nu .
\end{align}
Equations~\eqref{eq:fluctuations} can be written in terms of the d'Alembertian of $g_{\mu \nu}$ as
\begin{subequations}
\label{eq:acousticfluctuations}
\begin{align}
\Box_g \psi_1 - 4 \lambda \psi_1 = 0, \\
\Box_g \psi_2 = 0.
\end{align}
\end{subequations}
We see from eqs.~\eqref{eq:acousticfluctuations} that the fluctuations propagate on a curved metric, called the acoustic metric, which in this case is conformal to the background Minkowski space eq.~\eqref{eq:acousticmetric}. Note that in this derivation there was no low-momentum approximation needed in order to derive the acoustic metric.
\section{Relation to standard rBEC}
\label{sec:relStBEC}
The following section is devoted to the connection of the current work with the previous results on relativistic BEC as analogue models presented in ~\cite{Fagnocchi:2010sn}. As such it can be skipped being not strictly needed for the overall understanding of the rest of this work.

As noted earlier, eq.~\eqref{eq:master} is exactly the starting equation of the previous work on relativistic BEC by one of the authors~\cite{Fagnocchi:2010sn}, in which the acoustic metric felt by the perturbations of condensate was derived for the first time. This work also showed that this acoustic metric coincides with the one derived in ref.~\cite{Visser:2010xv} (see also~\cite{Bilic:1999sq, Moncrief} for an earlier derivation) for the relativistic flow of an inviscid, irrotational fluid with a barotropic equation of state. The perturbations of such a fluid propagate on an acoustic geometry which is disformally related to the background Minkowski space,
\begin{equation}\label{metr}
g_{\mu\nu}=\rho\frac{c}{c_{s}}\left[\eta_{\mu\nu}+\left(1-\frac{c_{s}^{2}}{c^{2}}\right)\frac{v_{\mu}v_{\nu}}{c^{2}}\right]\, ,
\end{equation},
 where $c_s$ is the speed of sound and $v^{\mu}=c u^{\mu}/\left\|u\right\|$ is the velocity of the fluid flow. Here $u^{\mu}\equiv \frac{\hbar}{m}\eta^{\mu\nu}\partial_{\nu}\theta$ is the usual four vector directly associated to the spacetime dependence of the phase of the background field written in the so called Madelung form $\varphi=\sqrt{\rho}e^{i\theta}$ (see ~\cite{Fagnocchi:2010sn} for a detailed discussion). This acoustic metric was later linked to previous studies of perturbations in the k-essence models~\cite{Bilic:2013qpa}. Given the disformal form of the acoustic metric found in all these studies, it might seem quite surprising that the acoustic metric in the present case is conformal to the flat space. More importantly, there is no Lorentz violation in the dynamics of the perturbations in our system: perturbations experience the same acoustic metric both at low and high momenta. Since the acoustic metric is conformally flat they propagate with the ``speed of sou
 nd" equal to $c$ (the speed of light).

The point is that we have assumed $\varphi_0$ to be real which is tantamount to have a constant phase $\theta$. {It is indeed possible to start from the general equations of ref.~\cite{Fagnocchi:2010sn} and try to see what happens in the limit in which the phase of the order parameter becomes a spacetime constant (in particular zero for simplicity). The results of this kind of limiting procedure are the following: first of all the dispersion relations in equation (38) of ref.~\cite{Fagnocchi:2010sn} becomes the dispersion relations for a massless and a massive mode that one can derive from eq.~\eqref{eq:fluctuations};  secondly, the parameter $b$ used in   ref.~\cite{Fagnocchi:2010sn} to define the low momentum limit, i.e., the approximation in which the acoustic metric can be derived, goes to infinity in the limit so that the low momentum limit is always satisfied. With the same limiting procedure it is also possible to show that the speed of sound becomes equal to the speed
  of light as it is in our current treatment and as should be expected by the dispersion relations that do not show anymore the Lorentz violating terms. Finally, another quantity that remain well define despite of the limit is the fluid four velocity, in fact one can easily see that $v^{\mu}v_{\mu}=-c^2$ and $v^{\mu}$ is finite.} This explanation has the weakness to not be straightforwardly applicable to massless particles, that we shall assume later, as the definition of $u^\mu$ becomes singular in this limit. But it seems to be possible to take the massless limit at the end of the calculation when no quantity directly depends on the mass.  The discussion of the massless boson gas condensation would need a separate treatment in case one wants to purse the fluid analogy.

{The previous discussion shows that the limiting procedure is well defined. The final step then is to see how the acoustic metric can be read off from the perturbation equations in such a limit of costant phase and if this metric is really conformally flat. For doing this is sufficient to start from equation (24) of~\cite{Fagnocchi:2010sn} and take the limit of constant phase, then one obtain $$(\Box+\eta^{\mu\nu}\partial_{\mu}\ln\rho\partial_{\nu})\psi-2\lambda\rho(\psi+\psi^{\dag})=0$$ that is equivalent to our~\eqref{eq:fluctuations}, where $\rho$ corresponds to $\varphi_{0}^{2}$. From this equation we already know that one can read out the conformally flat acoustic metric felt by the perturbations. Note also that the same conclusion can be obtained starting from eq.~\eqref{metr} and looking at the case in which $c_{s}=c$.}
It is important to note that the equality between the speed of sound and the speed of light gives rise to the fact that in this system there has no interpolation phase between two relativity groups with two limit speeds ($c_s$ in the IR limit, $c$ in the UV one) and the relativity group remains always the same at any energy. This is hence an example of a model of emergent space-time where the low and high energy regime share the same Lorentz invariance. This point is not trivial since, as far as we know, there is no toy model of emergent spacetime in which Lorentz violation is screened in this way at the lowest order of perturbation theory. The case at hand shows that it can be possible at the price of some non-trivial conditions on the background system.

While the above discussion shows how the current result is related to that of ref.~\cite{Fagnocchi:2010sn}, we should also stress that the starting formalism of the two works are indeed different. In the previous work the condensation was assumed {\em a priori} and the rest followed, here we have used the Grand Canonical formalism that is more suited to show that a condensation actually happens and also permits to derive the critical temperature for the interacting case. The crucial feature is that this formalism allowed  us to single out explicitly the chemical potential that gives to us a mass scale that we will use in the next sections in order to rescale the fields.

\section{Emergent Nordstr\"om gravity}
\label{sec:emergence}

In sec.~\ref{susbec:perturbations} we saw that the fluctuations of the condensate, also called the quasi-particle excitations, are oblivious of the flat background metric. They instead experience a curved geometry dictated by the condensate and the background. On the other hand, they back-react on the condensate through the relativistic generalization of the Gross-Pitaevskii equation~\eqref{eq:GP}. It is natural to ask if it is possible to have a geometric description of the dynamics of the condensate too. \par
The Ricci tensor of the acoustic metric~\eqref{eq:acousticmetric} can be
calculated to be
\begin{align}
\label{eq:Ricci}
R_g = - 6\, \frac{\Box\, \varphi_0}{\varphi_0^3}
\end{align}
Dividing the relativistic Gross-Pitaevskii equation by $\varphi_0^3$, eq.~\eqref{eq:GP} can be written as
\begin{align}\label{eq:massterm}
R_g +6\frac{m^2}{\varphi_{0}^2}+ 12\lambda = \langle T_{\rm qp} \rangle,
\end{align}
where we have defined $\langle T_{\rm qp} \rangle := -12 \lambda \left[ \,3\, \langle\psi_1^2\rangle + \langle\psi_2^2\rangle\, \right]$ and the subscript ``qp" reminds us that this quantity is determined by the quasi-particle excitations of the condensate.

Eq.~\eqref{eq:massterm} is evidently reminiscent of the Einstein--Fokker equation describing Nordstr\"om gravity~\cite{Deruelle:2011wu, Giulini:2006ry},
\begin{equation}
R+\Lambda=24\pi \frac{G_{\rm N}}{c^4}\, T,
\label{eq:realNord}
\end{equation}
where $R$ and $T$ are, respectively, the Ricci scalar and the trace of the stress-energy tensor of matter.
Unfortunately, the gravitational analogy of our equation is spoiled by the mass term. Therefore we will consider our system in the zero mass limit.
 Notice that, as discussed earlier, this limit does not spoil the presence of a condensate (see eq.~\eqref{eq:Tcrit}) or the uniqueness of the Lorentz group for constituents and excitations found in sec.~\ref{sec:relStBEC}. We shall come back to the physical reasons for this limit in the discussion section.
\par

The striking resemblance of equations \eqref{eq:massterm} with zero mass term and \eqref{eq:realNord} should not distract us from the need of one more step before comparing them. Indeed, the dimensions of the various quantities appearing in eq.~\eqref{eq:massterm} are not canonical and need to be fixed for such comparison to be meaningful. This is due to the fact that, as is usual in the analogue gravity literature, our acoustic metric is a dimensional quantity because $\varphi_0$ is dimensional. The fractional perturbations $\psi_1$ and $\psi_2$, on the other hand, are dimensionless.
We therefore need rescaling of the fields in order to have a dimensionless metric and (mass) dimension one scalar fields propagating on the curved metric.
We relegate the detailed discussion of these rescalings in appendix~\ref{app:redefinition}. The upshot of this dimensional analysis is that we need to scale the field $\varphi_0 \rightarrow \dfrac{\mu}{\sqrt{c\hbar}} \varphi_0$ and perturbation $\psi \rightarrow  \dfrac{\sqrt{c\hbar}}{\mu} \psi$. Finally, using these rescaled quantities we can rewrite  eq.~\eqref{eq:massterm} (with $m=0$) in the form of eq.~\eqref{eq:realNord} as
\begin{equation}
\label{eq:emergentNord}
R+\Lambda_{\rm eff} = \, \langle T_{\rm qp} \rangle,
\end{equation}
where $\Lambda_{eff}\equiv 12\lambda\frac{\mu^{2}}{c\hbar}$ and $T_{\rm qp}$ here and in the following is the same expression as in \eqref{eq:massterm} but with the mass dimension one fields.
Equations of motion of the quasi-particles \eqref{eq:fluctuations} can also be rewritten in terms of the rescaled fields as
\begin{subequations}
\label{eq:fluctuations-rescaled}
\begin{align}
\Box_g \psi_1 - \frac{4 \lambda \mu^2}{\hbar c }\psi_1 = 0, \\
\Box_g \psi_2 = 0,
\end{align}
\end{subequations}
where all quantities, including the $\Box_g$ operator, now pertain to those of the rescaled fields. \par

\subsection{Stress Energy Tensor and Newton constant}
\label{subsec:SET}
As a final step in order to verify the emergence of a true Nordstr\"om gravity theory from our system we still need to prove that $T_{\rm qp}$ is indeed related to the trace of the stress energy tensor for the analogue matter fields, i.e. our quasiparticles $T$. This turns out to be indeed the case and the proportionality factors relating these quantities will allow us to identify the effective Newton constant for this analogue system $G_{\rm eff}$. In the following we refer the reader mainly to appendix \ref{app:set} for technicalities and just state the main results.

We have seen before that perturbations feel an acoustic conformally flat metric~\eqref{eq:acousticmetric}, in appendix~\ref{app:action} is shown in detail how to write the effective action~\eqref{eq:Lagrangian} in a geometric form in terms of this acoustic metric (note that the equations for perturbations~\eqref{eq:fluctuations-rescaled} can also be derived from this effective action). Said that, we want to compute the stress energy tensor for the perturbations by varying the action with respect to the acoustic metric, i.e.
\begin{equation}
T_{\mu\nu}\equiv -\frac{1}{\sqrt{-g}}\frac{\delta\left(\sqrt{-g}\mathcal{S}_{2}\right)}{\delta g^{\mu\nu}},
\end{equation}
where $\mathcal{S}_{2}$ is only the quadratic (in perturbations) part of the action (see eq.~\eqref{s2}) as the linear part $\mathcal{S}_{1}$ is shown (see again appendix \ref{app:set}) to give no contribution to the trace of the stress energy tensor.

The final result for the expectation value of the trace of the stress-energy tensor in the background of $g_{\mu \nu} = \dfrac{\mu^2}{\hbar c} \varphi_0^2 \eta_{\mu \nu}$ is given by
\begin{equation}
\langle T\rangle =-2\lambda\frac{\mu^2}{c \hbar} \left[3\langle\psi_{1}^{2}\rangle+\langle\psi_{2}^{2}\rangle\right]=\frac{1}{6}\frac{\mu^2}{c\hbar}\langle T_{qp}\rangle.
\end{equation}
Due to this last expression one sees that the RHS of eq.~\eqref{eq:emergentNord} is actually given by $\frac{6c\hbar}{\mu^2}\langle T\rangle$ and hence our emergent Nordstr\"om gravity equation will be exactly of the form \eqref{eq:realNord} with the identification $G_{\rm eff}=\hbar c^5/(4\pi\mu^2)$. This value corresponds to an emergent analogue Planck scale $M_{\rm Pl}=\mu\sqrt{4\pi}/c^2$.

We have thus succeeded in expressing the dynamics of the background for our rBEC analogue model in a geometric language
\begin{equation}
\label{eq:emergentNordDef}
R+\Lambda_{\rm eff} = 24\pi \frac{G_{\rm eff}}{c^4}\, \langle T \rangle.
\end{equation}
The acoustic metric itself is sourced by the expectation value of the trace of the stress-energy tensor of the perturbations of the condensate playing the role of the matter.
These matter fields in turns propagate relativistically on a conformally flat acoustic metric~\eqref{eq:acousticmetric} with equations~\eqref{eq:fluctuations-rescaled}.

A final comment is deserved by the emergent, positive, cosmological constant term $\Lambda_{\rm eff}$.
The quantity of interest for what concern the usual cosmological constant problem is the ratio between the energy density associated to the (emergent) cosmological constant
$\epsilon_{\Lambda_{\rm eff}} \sim \left(\dfrac{\Lambda_{\rm eff} c^{4}}{G_{\rm eff}}\right)$ and the emergent Planck energy density $\epsilon_{\rm pl} \sim  \dfrac{c^{7}}{\hbar G_{\rm eff}^{2}}$.
In our case this ratio is given by
\begin{equation}
\label{eq:ratio}
	\frac{\epsilon_{\Lambda_{\rm eff}}}{\epsilon_{\rm pl}}\simeq \frac{3\lambda\hbar c}{\pi}.
\end{equation}
As one can see the ratio is proportional to $\lambda\hbar$ and so is clearly pretty small due to the presence of Planck constant and of the natural assumption of a weakly interacting system.
Of course in principle this term can be ``renormalised" by the vacuum contribution of the matter fields (basically the vacuum expectation value $\langle T \rangle$). It is however non-trivial, and beyond the scope of the present work, to split our ground state in a matter and vacuum part as it is not an eigenstate of the number operator (which in our relativistic system is not conserved).

\section{Summary and Discussion}
\label{sec:discuss}

In this paper we have studied the relativistic Bose--Einstein condensation in a theory of massless complex scalar field with a quartic coupling. Below the critical temperature the $U(1)$ symmetry is broken resulting in the non-zero value of the expectation value of the field --- the condensate. We showed that the dynamics of the condensate is described by the relativistic generalisation of the Gross--Pitaevskii equation given in eq.~\eqref{eq:GP}. The fluctuations of the condensate experience the presence of the condensate through the acoustic metric, eq.~\eqref{eq:acousticmetric} that, with the particular background state chosen here, turns out to be conformal to the flat Minkowski metric.
Propagation of the two components of the perturbation is described by eqs.~\eqref{eq:fluctuations-rescaled} which are just the Klein-Gordon equations for  massive and massless scalar fields on the curved background provided by the acoustic metric. Perturbations in turn gravitate through the trace of their stress-energy tensor that is calculated in detail in the appendix~\ref{app:set}. The dynamics of the acoustic metric is governed by the analogue Einstein--Fokker equation~\eqref{eq:emergentNord}, which is the equation of motion for the Nordstr\"om gravity with cosmological constant. To the best of our knowledge this is the first study of the emergence of Lorentz invariant dynamics for the emergent spacetime in an analogue model (see however ref.~\cite{Gielen:2013naa}). As a side remark, note also that the emergence of only conformally flat analogue spacetimes is in no way a trivial result since cosmological solutions in GR are conformally flat as well and nevertheless they i
 ncorporate characteristic features like expansion of the Universe and cosmological particle creation.

The central assumption that has permitted us to carry out the geometrical interpretation of the model is the reality of the order parameter. Thanks to this it was possible to have a conformally flat acoustic metric and to rewrite the background equation in a geometrical form. In the general case in which the order parameter is complex there is does not seem to be much hope to cast the non-linear Klein-Gordon equation for the background in a geometrical form, although an acoustic metric can still be derived and is in general a disformal metric~\eqref{metr}. This is due to the fact that the general disformal acoustic metric depends both on the (derivative of) phase and the modulus of the order parameter but the background equation is too simple to describe the dynamics of both the (derivative of) phase and the modulus of the order parameter, so cannot be recast in a background independent form. The reality of the condensate, on the other hand, leaves only one degree of freedom 
 to play with and hence at best one can only hope to recover a scalar theory of gravity such as the Nordstr\"om one in this limit. It would be interesting to further characterise the particular background state that has to be chosen in order to recover a gravitational dynamics.

Another necessary approximation for the emerging Nordstr\"om gravity is the zero mass limit of the underlying atoms.
 Although we have seen that the massless case is not pathological from the point of view of Bose--Einstein condensation, one should be aware that strictly speaking such a limit is not necessary since it is sufficient to require for the mass term in eq.~\eqref{eq:massterm} to be negligible with respect to the others (though this would call for a careful analysis and is beyond the scope of this work). One might wonder why the mass term ruins the geometrical interpretation of the equation. Let us just notice that this term breaks the conformal invariance of the background equation~\eqref{eq:GP}. Similarly, the addition of higher order interactions (see also discussion below) would break the conformal invariance of~\eqref{eq:GP} and spoil the possibility to recast the equation in a geometric form. It would be interesting to further investigate this apparent link and pinpoint the exact connection (if any) between conformal invariance of the background equation and its viability for a geometric interpretation.

From a pure EFT point of view it is clear that other interaction terms are admissible and, as mentioned above, higher mass dimension interaction terms, $\phi^n$ as well as a cubic term (which could be discarded anyway by parity arguments), would end up spoiling the geometrical interpretation of the theory.
However, while in principle the aforementioned higher order interactions are allowed, there are good physical reasons for the $\lambda \phi^4$ interaction to be the most relevant one. In fact, such term models two body interactions which are generically dominant in dilute systems as the condensate that we have considered here. Higher order interaction terms will not only be irrelevant from an EFT point of view but will be associated to many-body interactions which will be generically subdominant.

 It is also interesting that we obtain quite naturally a cosmological constant term whose size is set by the coupling constant $\lambda$ and the chemical potential $\mu$. Remarkably, the emergent cosmological constant is such that the ratio between its energy density and the energy density associated to the emergent Planck length eq.~\eqref{eq:ratio} is small: so there  is no ``cosmological constant problem" (in the sense of unnatural smallness) present in such emergent gravity systems. This result is in close analogy with the non-relativistic case discussed in ref.~\cite{Finazzi:2011zw}.
It is however important to stress that in our relativistic case the recovery of  such a term is strongly dependent on the choice of the particular interaction term characterising the initial Lagrangian \eqref{eq:Lagrangiana}, ie. the $\lambda\phi^4$ one and it is not present in the non-relativistic case. 

Indeed, as discussed in \cite{Finazzi:2011zw}, the small, negative, cosmological constant term found in the non-relativistic BEC is basically due to the depletion factor, i.e. to that ever present atoms which are not in the condensate phase. This is a pure quantum effect due to the quantum inequivalence of the phonon and atomic vacua.  The relativistic case shows instead a ``bare" gravitational constant term, simply stemming from the $\phi^4$ term, which is there independently from the vacuum expectation value $\langle T \rangle$ contribution (the relativistic generalisation of the term associated to depletion in the non-relativistic BEC). Of course one can recover the non-relativistic BEC case from the relativistic BEC (see \cite{Fagnocchi:2010sn}). In this case the dimensional bare coupling constant ($\Lambda_{\rm eff}=12\lambda\mu^2/c\hbar$, see appendix A) goes to zero as $c\rightarrow\infty$ and only the ``depletion" contribution will remain.

Finally, Nordstr\"om gravity is only a scalar theory of gravity and has been falsified by experiments, for example, it does not predict the bending of light. However, it is the only other known theory in 4 dim that satisfies the strong equivalence principle~\cite{DiCasola:2013yga}. With the aim of getting closer to emerge General Relativity, one necessarily needs to look for richer Lagrangians than that in eq.~\eqref{eq:Lagrangiana}. Of course, emergence of a theory characterised by spin-2 graviton would open the door to a possible conflict with the Weinberg--Witten theorem~\cite{20623}. However, one may guess that analogue models (or analogue model inspired systems) will generically lead to Lagrangians which show Lorentz invariance and background independence only as approximate symmetries for the lowest order in the perturbative expansion.
The relativistic model proposed here shows that, at least at the level of linear perturbations, such symmetries are realised both in the equations of the linear perturbations as well as in those describing the dynamics of the background. As such it might serve as toy model for the use of emergent gravity  scenarios in investigating, e.g. geometrogenesis (here the condensation process)~\cite{Oriti:2013jga} or  the nature of spacetime singularities in this framework. We hope to come back to these and related issues in the near future.

\section{Acknowledgments}
The authors are grateful to Stefano Finazzi and Lorenzo Sindoni for illuminating discussions and useful comments on the manuscript. We also wish to thank Dionigi Benincasa, Eolo Di Casola, Andrea Gambassi and Andrea Trombettoni for useful discussions. AM thanks Joseph Kapusta for correspondence related to the Gross--Pit{\ae}vskii equation.

\appendix
\section{Field redefinition}
\label{app:redefinition}
Here we are going to redefine the fields in such a way to have a dimensionless acoustic metric and mass dimension one scalar fields propagating on it. In order to do so let us do a little bit of dimensional analysis. By looking to the standard kinetic term for a scalar field in 4D one has that the dimension of the field is given by $$[\phi]=\sqrt{\frac{ML}{T^2}},$$ in accordance with the fact that the mass dimension is one in 4D. The chemical potential has the dimension of an energy and so $$[\mu]=\frac{ML^2}{T^2}.$$ Since we have an interaction term of the form $\lambda\phi^4$ we have also $$[\lambda]=\frac{T^2}{ML^3}.$$ First of all we want to redefine the background field (the condensate part) $\phi_{0}$ in such a way to render it dimensionless, this can be achived by the following redefinition $$\tilde{\varphi}_{0}= \frac{\sqrt{\hbar c}}{\mu}\varphi_{0},$$ and this is the only way given the fact that we have only one mass scale given by the chemical potential (that has ma
 ss dimension one)
In analogy we have to redefine the perturbation field in the following way $$\tilde{\psi}=\frac{\mu}{\sqrt{\hbar c}}\psi.$$ Now we have a new acoustic metric given by $$\tilde{g}_{\mu\nu}=\tilde{\varphi}_{0}^{2}\eta_{\mu\nu},$$ in term of this the perturbation equations became $$\Box_{\tilde{g}}\tilde{\psi}_{1}-4\lambda\frac{\mu^{2}}{\hbar c}\tilde{\psi}_{1}=0,$$ $$\Box_{\tilde{g}}\tilde{\psi}_{2}=0.$$ The background equation instead become $$\tilde{R}+12\lambda\frac{\mu^{2}}{c \hbar}=0,$$ and so we can call cosmological constant the factor $\Lambda_{\rm eff}\equiv 12\lambda\frac{\mu^{2}}{c\hbar}$ that has in fact the right dimension, $1/L^{2}$. Now from only dimensional arguments it is easy to guess what will be the emergent gravitational constant in our model, in fact the only combination of constants of the model with the right dimension is $$\frac{\hbar c^{5}}{\mu^{2}}\equiv G,$$ and so the would be Planck mass is dimensionally set by $\frac{\mu}{c^{2}}$.
\section{Action in geometrical form}
\label{app:action}
In this Appendix we are going to rewrite the action for background field and perturbations making explicit use of the acoustic metric. We will do it with the non-redefined field and in natural units for the moment. In order to do so we have to rewrite the effective Lagrangian~\eqref{eq:Lagrangian}, after get rid of the $\mu$ dependent term, splitting the background field and the fractional perturbation, $\phi=\varphi_{0}(1+\psi)$ (here and in the following, for economy of space we will not split the perturbations in real and imaginary part unless needed). In this way one obtain the following
\begin{equation}
\mathcal{L}_{\rm eff}=\mathcal{L}_{0}+\mathcal{L}_{1}+\mathcal{L}_{2}+\mathcal{L}_{3,4},
\end{equation}
where the number in the end represent the number of the perturbation fields in the Lagrangians and
\begin{align}\label{eq:lag}
& \mathcal{L}_{0}=-\eta^{\mu\nu}\partial_{\mu}\varphi_{0}\partial_{\nu}\varphi_{0}-m^{2}\varphi_{0}^{2}-\lambda\varphi_{0}^{4} \\
& \mathcal{L}_{1}= \left(-\eta^{\mu\nu}\partial_{\mu}\varphi_{0}\partial_{\nu}\varphi_{0}-m^{2}\varphi_{0}^{2}-2\lambda\varphi_{0}^{4}\right)(\psi^{*}+\psi)-\eta^{\mu\nu}\partial_{\mu}\varphi_{0}\varphi_{0}\partial_{\nu}\psi-\eta^{\mu\nu}\partial_{\mu}\varphi_{0}\varphi_{0}\partial_{\nu}\psi^{*}
\end{align}
\begin{align}
&\mathcal{L}_{2} = \left(-\eta^{\mu\nu}\partial_{\mu}\varphi_{0}\partial_{\nu}\varphi_{0}-m^{2}\varphi_{0}^{2}\right)(\psi^{*}\psi)-\lambda\varphi_{0}^{4}\left(\psi\psi+\psi^{*}\psi^{*}+4\psi^{*}\psi\right)-\eta^{\mu\nu}\varphi_{0}^{2}\partial_{\mu}\psi^{*}\partial_{\nu}\psi \\ \nonumber
& -\eta^{\mu\nu}\varphi_{0}\partial_{\mu}\varphi_{0} \psi^{*}\partial_{\nu}\psi-\eta^{\mu\nu}\partial_{\mu}\psi^{*} \varphi_{0}\partial_{\nu}\varphi_{0} \psi \\
& \mathcal{L}_{3,4}= -\lambda\varphi_{0}^{4}\left(2\psi^{*}\psi\psi+2\psi^{*}\psi^{*}\psi+\psi^{*}\psi^{*}\psi\psi\right)
\end{align}

Now we are going to put the action of the theory, up to quadratic terms in the perturbation, in a geometrical form. In order to do so we will integrate by parts terms in the above Lagrangian ignoring the boundary terms that will arise.
\\
First of all remember that for us
$$
g_{\mu\nu}=\varphi_{0}^{2}\eta_{\mu\nu},
$$
$$
\sqrt{-g}=\varphi_{0}^{4}.
$$
Then the expression
$$
-\eta^{\mu\nu}\varphi_{0}\partial_{\mu}\varphi_{0}\partial_{\mu}\psi-\eta^{\mu\nu}\varphi_{0}\partial_{\mu}\varphi_{0}\partial_{\mu}\psi^{*}-\eta^{\mu\nu}\partial_{\mu}\varphi_{0}\partial_{\mu}\varphi_{0}(\psi^{*}+\psi)
$$
after integration by part of the first two terms become
\begin{align}
& \eta^{\mu\nu}\partial_{\nu}\varphi_{0}\partial_{\mu}\varphi_{0}\psi+\eta^{\mu\nu}\varphi_{0}\partial_{\nu}\partial_{\mu}\varphi_{0}\psi+\eta^{\mu\nu}\partial_{\nu}\varphi_{0}\partial_{\mu}\varphi_{0}\psi^{*}\\ \nonumber
& +\eta^{\mu\nu}\varphi_{0}\partial_{\nu}\partial_{\mu}\varphi_{0}\psi^{*}-\eta^{\mu\nu}\partial_{\mu}\varphi_{0}\partial_{\mu}\varphi_{0}(\psi^{*}+\psi)=\eta^{\mu\nu}\varphi_{0}\partial_{\nu}\partial_{\mu}\varphi_{0}\psi+\eta^{\mu\nu}\varphi_{0}\partial_{\nu}\partial_{\mu}\varphi_{0}\psi^{*}
\end{align}
Now lets look at the term in the action
\begin{equation}
\int d^{4}x\varphi_{0}\Box\varphi_{0}(\psi+\psi^{*})=\int d^{4}x\sqrt{-g}\frac{\varphi_{0}\Box\varphi_{0}}{\varphi_{0}^{4}}\frac{-6}{-6}(\psi+\psi^{*})=\frac{-1}{6}\int\sqrt{-g}R(\psi+\psi^{*})
\end{equation}
Let us now pass to the other terms and proceed in the same way as above. The term
$$
-\eta^{\mu\nu}\partial_{\mu}\varphi_{0}\partial_{\nu}\varphi_{0}\psi^{*}\psi-\eta^{\mu\nu}\partial_{\mu}\varphi_{0}\varphi_{0}\psi^{*}\partial_{\nu}\psi-\eta^{\mu\nu}\partial_{\mu}\varphi_{0}\varphi_{0}\partial_{\nu}\psi^{*}\
\psi
$$
after integration by part of the first term become
\begin{align}
& \eta^{\mu\nu}\varphi_{0}\partial_{\mu}\partial_{\nu}\varphi_{0}\psi^{*}\psi+\eta^{\mu\nu}\varphi_{0}\partial_{\nu}\varphi_{0}\partial_{\mu}\psi^{*}\psi+\eta^{\mu\nu}\varphi_{0}\partial_{\nu}\varphi_{0}\psi^{*}\partial_{\mu}\psi\\ \nonumber
& -\eta^{\mu\nu}\partial_{\mu}\varphi_{0}\varphi_{0}\psi^{*}\partial_{\nu}\psi-\eta^{\mu\nu}\partial_{\mu}\varphi_{0}\varphi_{0}\partial_{\nu}\psi^{*}\psi=\eta^{\mu\nu}\varphi_{0}\partial_{\mu}\varphi_{0}\partial_{\nu}\varphi_{0}(\psi^{*}\psi),
\end{align}
and so in the action
\begin{equation}
\int d^{4}x\eta^{\mu\nu}\varphi_{0}\partial_{\mu}\varphi_{0}\partial_{\nu}\varphi_{0}(\psi^{*}\psi)=-\frac{1}{6}\int d^{4}x\sqrt{-g}R(\psi^{*}\psi)
\end{equation}
$$
-\eta^{\mu\nu}\partial_{\mu}\varphi_{0}\partial_{\nu}\varphi_{0}\rightarrow \eta^{\mu\nu}\varphi_{0}\partial_{\mu}\partial_{\nu}\varphi_{0}\rightarrow\int d^{4}x\sqrt{-g}\frac{\varphi_{0}\Box\varphi_{0}}{\varphi_{0}^{4}}=-\frac{1}{6}\int d^{4}x\sqrt{-g}R.
$$
We have now other two remaining terms for which we do not need to integrate by part. The first one is
\begin{align}
& -m^{2}\varphi_{0}^{2}\left[1+\psi^{*}+\psi+\psi^{*}\psi\right]\\ \nonumber
& -\lambda\varphi_{0}^{4}\left[1+2(\psi^{*}+\psi)+\psi\psi+\psi^{*}\psi^{*}+4\psi^{*}\psi\right]
\end{align}
that become in the action
\begin{align}
& -\int d^{4}x\sqrt{-g}\left\{m^{2}\varphi_{0}^{-2}\left[1+\psi^{*}+\psi+\psi^{*}\psi\right]\right.\\ \nonumber
& \left.+\lambda\left[1+2(\psi^{*}+\psi)+\psi\psi+\psi^{*}\psi^{*}+4\psi^{*}\psi\right]\right\}.
\end{align}
The second and last term we are left with is
$$
-\eta^{\mu\nu}\varphi_{0}^{2}\partial_{\mu}\psi^{*}\partial_{\nu}\psi\rightarrow-\int d^{4}x\sqrt{-g}\frac{\varphi_{0}^{2}\eta^{\mu\nu}}{\varphi_{0}^{4}}\partial_{\mu}\psi^{*}\partial_{\nu}\psi=-\int d^{4}x\sqrt{-g}g^{\mu\nu}\partial_{\mu}\psi^{*}\partial_{\nu}\psi.
$$
Putting all together, and also putting the mass to be zero, we have the following action
\begin{align}\label{eq:EndAct}
S & =\int d^{4}x\sqrt{-g}\left\{-\frac{1}{6}R+\frac{-1}{6}R(\psi+\psi^{*})-\frac{1}{6}R(\psi^{*}\psi)\right. \\ \nonumber
& \left. -\lambda\left[1+2(\psi^{*}+\psi)+\psi\psi+ \psi^{*}\psi^{*}+4\psi^{*}\psi\right]-g^{\mu\nu}\partial_{\mu}\psi^{*}\partial_{\nu}\psi\frac{}{}\right\}.
\end{align}
\section{Stress-Energy tensor}
\label{app:set}
In this last appendix we will report the detailed calculation for the stress energy tensor and its trace. So we want to calculate
\begin{equation}
T_{\mu\nu}\equiv -\frac{1}{\sqrt{-g}}\frac{\delta\left(\sqrt{-g}\mathcal{L}_{2}\right)}{\delta g^{\mu\nu}}.
\end{equation}
We will consider the quadratic part of the action in the perturbations fields given by (we will also use the redefined quantities omitting the tilde)
\begin{align}\label{s2}
\mathcal{S}_{2}&\equiv \frac{1}{c}\int d^{4}x \sqrt{-g}\mathcal{L}^{geom}_{2} \nonumber \\ &= -\int d^{4}x\sqrt{-g}\left\{\frac{1}{6}R(\psi^{*}\psi)+\frac{1}{12}\Lambda\left[\psi\psi+ \psi^{*}\psi^{*}+4\psi^{*}\psi\right]+g^{\mu\nu}\partial_{\mu}\psi^{*}\partial_{\nu}\psi\right\}.
\end{align}
In the end we will also show that ideed the linear part of the action in the perturbations gives no contribution to the SET.
We will also need the following relations
\begin{subequations}
\label{eq:rel}
\begin{align}
\delta(\sqrt{-g})=-\frac{1}{2}\sqrt{-g}g_{\mu\nu}\delta g^{\mu\nu}, \\
\delta R=R_{\mu\nu}\delta g^{\mu\nu}+g_{\mu\nu}\Box_{g}\delta g^{\mu\nu}-\nabla_{\mu}\nabla_{\nu}\delta g^{\mu\nu}, \\
\int d^{4}x\sqrt{-g}[f\delta R]=\int d^{4}x\sqrt{-g}[fR_{\mu\nu}+g_{\mu\nu}\Box_{g} f-\nabla_{\mu}\nabla_{\nu}f]\delta g^{\mu\nu},
\end{align}
\end{subequations}
where the third one follow from the second integrating by parts and neglecting boundary terms. Then we have
\begin{align}
\delta\mathcal{S}_{2}= & -\frac{1}{c}\int d^{4}x\sqrt{-g}\left\{\frac{1}{6}R_{\mu\nu}\psi^{*}\psi+\frac{1}{6}g_{\mu\nu}\Box_{g}\psi^{*}\psi+\frac{2}{6}g_{\mu\nu}\nabla^{a}\psi^{*}\nabla_{a}\psi+\frac{1}{6}g_{\mu\nu}\psi^{*}\Box_{g}\psi\right. \\ \nonumber
& \left.-\frac{1}{6}\nabla_{\mu}\nabla_{\nu}\psi^{*}\psi-\frac{1}{6}\nabla_{\nu}\psi^{*}\nabla_{\mu}\psi-\frac{2}{6}\nabla_{\mu}\psi^{*}\nabla_{\nu}\psi \right.\\ \nonumber
& \left. -\frac{\Lambda}{12}\frac{1}{2}g_{\mu\nu}\left[\psi\psi+ \psi^{*}\psi^{*}+4\psi^{*}\psi\right]-\frac{1}{2}g_{\mu\nu}\partial_{\alpha}\psi^{*}\partial^{\alpha}\psi+\partial_{\mu}\psi^{*}\partial^{\nu}\psi-\frac{1}{6}R\frac{1}{2}g_{\mu\nu}\psi^{*}\psi\right\}\delta g^{\mu\nu}.
\end{align}
Then the stress energy tensor is simply given by
\begin{align}
T_{\mu\nu}= & \frac{1}{6}G_{\mu\nu}\psi^{*}\psi+\frac{1}{6}g_{\mu\nu}\Box_{g}\psi^{*}\psi+\frac{2}{6}g_{\mu\nu}\nabla^{a}\psi^{*}\nabla_{a}\psi+\frac{1}{6}g_{\mu\nu}\psi^{*}\Box_{g}\psi \\ \nonumber
& -\frac{1}{6}\nabla_{\mu}\nabla_{\nu}\psi^{*}\psi-\frac{1}{6}\nabla_{\nu}\psi^{*}\nabla_{\mu}\psi-\frac{2}{6}\nabla_{\mu}\psi^{*}\nabla_{\nu}\psi \\ \nonumber
& -\frac{\Lambda}{12}\frac{1}{2}g_{\mu\nu}\left[\psi\psi+ \psi^{*}\psi^{*}+4\psi^{*}\psi\right]-\frac{1}{2}g_{\mu\nu}\partial_{\alpha}\psi^{*}\partial^{\alpha}\psi+\partial_{\mu}\psi^{*}\partial^{\nu}\psi,
\end{align}
and its trace is given by
\begin{align}
T=& -\left(\frac{R+\Lambda}{6}\right)\psi^{*}\psi-\frac{\Lambda}{6}\left[\psi\psi+\psi^{*}\psi^{*}+3\psi^{*}\psi\right] \\ \nonumber
& +\Box_{g}\psi^{*}\psi\left(\frac{2}{3}-\frac{1}{6}\right)+\psi^{*}\Box_{g}\psi\left(\frac{2}{3}-\frac{1}{6}\right)+\partial_{\alpha}\psi^{*}\partial^{\alpha}\psi\left(-1-\frac{1}{3}+\frac{4}{3}\right).
\end{align}
Finally, using the background and the perturbations equations
\begin{subequations}
\begin{align}
R+\Lambda =0, \\
\Box_{g}\psi =\frac{\Lambda}{6}\left(\psi+\psi^{*}\right),
\end{align}
\end{subequations}
and splitting the field in imaginary and real part, we end up with
\begin{equation}
\label{eq:traceSET}
T=-2\lambda\frac{\mu^{2}}{c\hbar}\left[3\psi_{1}^{2}+\psi_{2}^{2}\right].
\end{equation}
To conclude this appendix we have to show that, as anticipated, the linear (in the perturbations) part of the action gives no contribution to the stress tensor. The linear part is given by
\begin{equation}
\mathcal{S}_{1}\propto-\int d^{4}x \sqrt{-g}\left\{\frac{1}{6}R(\psi^{*}+\psi)+\frac{1}{6}\Lambda(\psi+ \psi^{*})\right\},
\end{equation}
so then following the same steps as before we have
\begin{align}
\delta\mathcal{S}_{1}\propto & -\int d^{4}x\sqrt{-g}\left\{\frac{1}{6}R_{\mu\nu}(\psi^{*}+\psi)+\frac{1}{6}g_{\mu\nu}(\Box_{g}\psi^{*}+\Box_{g}\psi)-\frac{1}{6}\left(\nabla_{\mu}\nabla_{\nu}\psi^{*}+\nabla_{\mu}\nabla_{\nu}\psi\right)\right. \\ \nonumber
& \left.-\frac{1}{6}R\frac{1}{2}(\psi^{*}+\psi)-\frac{\Lambda}{6}(\psi^{*}+\psi)\frac{1}{2}g_{\mu\nu}\right\}\delta g^{\mu\nu}.
\end{align}
Now is easy to see what is the contribution to the trace of the stress energy tensor given by the linear term
\begin{equation}
T^{(1)}=-\left(\frac{R+\Lambda}{6}\right)(\psi+\psi^{*})+\frac{1}{2}\left(\Box_{g}\psi^{*}+\Box_{g}\psi-\frac{2\Lambda}{6}(\psi+\psi^{*})\right),
\end{equation}
and using the background and perturbations equations this give zero.
\bibliography{relBEC_refs}

\begin{thebibliography}{47}%
\makeatletter
\providecommand \@ifxundefined [1]{%
 \@ifx{#1\undefined}
}%
\providecommand \@ifnum [1]{%
 \ifnum #1\expandafter \@firstoftwo
 \else \expandafter \@secondoftwo
 \fi
}%
\providecommand \@ifx [1]{%
 \ifx #1\expandafter \@firstoftwo
 \else \expandafter \@secondoftwo
 \fi
}%
\providecommand \natexlab [1]{#1}%
\providecommand \enquote  [1]{``#1''}%
\providecommand \bibnamefont  [1]{#1}%
\providecommand \bibfnamefont [1]{#1}%
\providecommand \citenamefont [1]{#1}%
\providecommand \href@noop [0]{\@secondoftwo}%
\providecommand \href [0]{\begingroup \@sanitize@url \@href}%
\providecommand \@href[1]{\@@startlink{#1}\@@href}%
\providecommand \@@href[1]{\endgroup#1\@@endlink}%
\providecommand \@sanitize@url [0]{\catcode `\\12\catcode `\$12\catcode
  `\&12\catcode `\#12\catcode `\^12\catcode `\_12\catcode `\%12\relax}%
\providecommand \@@startlink[1]{}%
\providecommand \@@endlink[0]{}%
\providecommand \url  [0]{\begingroup\@sanitize@url \@url }%
\providecommand \@url [1]{\endgroup\@href {#1}{\urlprefix }}%
\providecommand \urlprefix  [0]{URL }%
\providecommand \Eprint [0]{\href }%
\providecommand \doibase [0]{http://dx.doi.org/}%
\providecommand \selectlanguage [0]{\@gobble}%
\providecommand \bibinfo  [0]{\@secondoftwo}%
\providecommand \bibfield  [0]{\@secondoftwo}%
\providecommand \translation [1]{[#1]}%
\providecommand \BibitemOpen [0]{}%
\providecommand \bibitemStop [0]{}%
\providecommand \bibitemNoStop [0]{.\EOS\space}%
\providecommand \EOS [0]{\spacefactor3000\relax}%
\providecommand \BibitemShut  [1]{\csname bibitem#1\endcsname}%
\let\auto@bib@innerbib\@empty
\bibitem [{\citenamefont {Bardeen}\ \emph {et~al.}(1973)\citenamefont
  {Bardeen}, \citenamefont {Carter},\ and\ \citenamefont
  {Hawking}}]{Bardeen:1973gs}%
  \BibitemOpen
  \bibfield  {author} {\bibinfo {author} {\bibfnamefont {J.~M.}\ \bibnamefont
  {Bardeen}}, \bibinfo {author} {\bibfnamefont {B.}~\bibnamefont {Carter}}, \
  and\ \bibinfo {author} {\bibfnamefont {S.}~\bibnamefont {Hawking}},\ }\href
  {\doibase 10.1007/BF01645742} {\bibfield  {journal} {\bibinfo  {journal}
  {Commun.Math.Phys.}\ }\textbf {\bibinfo {volume} {31}},\ \bibinfo {pages}
  {161} (\bibinfo {year} {1973})}\BibitemShut {NoStop}%
\bibitem [{\citenamefont {Hawking}(1975)}]{Hawking:1974sw}%
  \BibitemOpen
  \bibfield  {author} {\bibinfo {author} {\bibfnamefont {S.}~\bibnamefont
  {Hawking}},\ }\href {\doibase 10.1007/BF02345020} {\bibfield  {journal}
  {\bibinfo  {journal} {Commun.Math.Phys.}\ }\textbf {\bibinfo {volume} {43}},\
  \bibinfo {pages} {199} (\bibinfo {year} {1975})}\BibitemShut {NoStop}%
\bibitem [{\citenamefont {Hawking}(1974)}]{Hawking:1974rv}%
  \BibitemOpen
  \bibfield  {author} {\bibinfo {author} {\bibfnamefont {S.}~\bibnamefont
  {Hawking}},\ }\href {\doibase 10.1038/248030a0} {\bibfield  {journal}
  {\bibinfo  {journal} {Nature}\ }\textbf {\bibinfo {volume} {248}},\ \bibinfo
  {pages} {30} (\bibinfo {year} {1974})}\BibitemShut {NoStop}%
\bibitem [{\citenamefont {Jacobson}(1995)}]{Jacobson:1995ab}%
  \BibitemOpen
  \bibfield  {author} {\bibinfo {author} {\bibfnamefont {T.}~\bibnamefont
  {Jacobson}},\ }\href {\doibase 10.1103/PhysRevLett.75.1260} {\bibfield
  {journal} {\bibinfo  {journal} {Phys.Rev.Lett.}\ }\textbf {\bibinfo {volume}
  {75}},\ \bibinfo {pages} {1260} (\bibinfo {year} {1995})},\ \Eprint
  {http://arxiv.org/abs/gr-qc/9504004} {arXiv:gr-qc/9504004 [gr-qc]}
  \BibitemShut {NoStop}%
\bibitem [{\citenamefont {Chirco}\ \emph {et~al.}(2014)\citenamefont {Chirco},
  \citenamefont {Haggard}, \citenamefont {Riello},\ and\ \citenamefont
  {Rovelli}}]{Chirco:2014saa}%
  \BibitemOpen
  \bibfield  {author} {\bibinfo {author} {\bibfnamefont {G.}~\bibnamefont
  {Chirco}}, \bibinfo {author} {\bibfnamefont {H.~M.}\ \bibnamefont {Haggard}},
  \bibinfo {author} {\bibfnamefont {A.}~\bibnamefont {Riello}}, \ and\ \bibinfo
  {author} {\bibfnamefont {C.}~\bibnamefont {Rovelli}},\ }\href@noop {} {\
  (\bibinfo {year} {2014})},\ \Eprint {http://arxiv.org/abs/1401.5262}
  {arXiv:1401.5262 [gr-qc]} \BibitemShut {NoStop}%
\bibitem [{\citenamefont {Thorne}\ \emph {et~al.}(1986)\citenamefont {Thorne},
  \citenamefont {Price},\ and\ \citenamefont {Macdonald}}]{Thorne:1986iy}%
  \BibitemOpen
  \bibfield  {author} {\bibinfo {author} {\bibfnamefont {K.~S.}\ \bibnamefont
  {Thorne}}, \bibinfo {author} {\bibfnamefont {R.}~\bibnamefont {Price}}, \
  and\ \bibinfo {author} {\bibfnamefont {D.}~\bibnamefont {Macdonald}},\
  }\href@noop {} {\emph {\bibinfo {title} {{Black holes: the Membrane
  paradigm}}}},\ edited by\ \bibinfo {editor} {\bibfnamefont {K.~S.}\
  \bibnamefont {Thorne}}\ (\bibinfo {year} {1986})\BibitemShut {NoStop}%
\bibitem [{\citenamefont {Damour}(1978)}]{Damour:1978cg}%
  \BibitemOpen
  \bibfield  {author} {\bibinfo {author} {\bibfnamefont {T.}~\bibnamefont
  {Damour}},\ }\href {\doibase 10.1103/PhysRevD.18.3598} {\bibfield  {journal}
  {\bibinfo  {journal} {Phys.Rev.}\ }\textbf {\bibinfo {volume} {D18}},\
  \bibinfo {pages} {3598} (\bibinfo {year} {1978})}\BibitemShut {NoStop}%
\bibitem [{\citenamefont {Sindoni}(2012)}]{Sindoni:2011ej}%
  \BibitemOpen
  \bibfield  {author} {\bibinfo {author} {\bibfnamefont {L.}~\bibnamefont
  {Sindoni}},\ }\href {\doibase 10.3842/SIGMA.2012.027} {\bibfield  {journal}
  {\bibinfo  {journal} {SIGMA}\ }\textbf {\bibinfo {volume} {8}},\ \bibinfo
  {pages} {027} (\bibinfo {year} {2012})},\ \Eprint
  {http://arxiv.org/abs/1110.0686} {arXiv:1110.0686 [gr-qc]} \BibitemShut
  {NoStop}%
\bibitem [{\citenamefont {Gielen}\ \emph {et~al.}(2013)\citenamefont {Gielen},
  \citenamefont {Oriti},\ and\ \citenamefont {Sindoni}}]{Gielen:2013naa}%
  \BibitemOpen
  \bibfield  {author} {\bibinfo {author} {\bibfnamefont {S.}~\bibnamefont
  {Gielen}}, \bibinfo {author} {\bibfnamefont {D.}~\bibnamefont {Oriti}}, \
  and\ \bibinfo {author} {\bibfnamefont {L.}~\bibnamefont {Sindoni}},\
  }\href@noop {} {\  (\bibinfo {year} {2013})},\ \Eprint
  {http://arxiv.org/abs/1311.1238} {arXiv:1311.1238 [gr-qc]} \BibitemShut
  {NoStop}%
\bibitem [{\citenamefont {Oriti}(2007)}]{Oriti:2007qd}%
  \BibitemOpen
  \bibfield  {author} {\bibinfo {author} {\bibfnamefont {D.}~\bibnamefont
  {Oriti}},\ }\href@noop {} {\bibfield  {journal} {\bibinfo  {journal} {PoS}\
  }\textbf {\bibinfo {volume} {QG-PH}},\ \bibinfo {pages} {030} (\bibinfo
  {year} {2007})},\ \Eprint {http://arxiv.org/abs/0710.3276} {arXiv:0710.3276
  [gr-qc]} \BibitemShut {NoStop}%
\bibitem [{\citenamefont {Hu}(2005)}]{Hu:2005ub}%
  \BibitemOpen
  \bibfield  {author} {\bibinfo {author} {\bibfnamefont {B.}~\bibnamefont
  {Hu}},\ }\href {\doibase 10.1007/s10773-005-8895-0} {\bibfield  {journal}
  {\bibinfo  {journal} {Int.J.Theor.Phys.}\ }\textbf {\bibinfo {volume} {44}},\
  \bibinfo {pages} {1785} (\bibinfo {year} {2005})},\ \Eprint
  {http://arxiv.org/abs/gr-qc/0503067} {arXiv:gr-qc/0503067 [gr-qc]}
  \BibitemShut {NoStop}%
\bibitem [{\citenamefont {Dreyer}(2007)}]{Dreyer:2007ws}%
  \BibitemOpen
  \bibfield  {author} {\bibinfo {author} {\bibfnamefont {O.}~\bibnamefont
  {Dreyer}},\ }\href@noop {} {\bibfield  {journal} {\bibinfo  {journal} {PoS}\
  }\textbf {\bibinfo {volume} {QG-PH}},\ \bibinfo {pages} {016} (\bibinfo
  {year} {2007})},\ \Eprint {http://arxiv.org/abs/0710.4350} {arXiv:0710.4350
  [gr-qc]} \BibitemShut {NoStop}%
\bibitem [{\citenamefont {Dreyer}(2006)}]{Dreyer:2006pp}%
  \BibitemOpen
  \bibfield  {author} {\bibinfo {author} {\bibfnamefont {O.}~\bibnamefont
  {Dreyer}},\ }\href@noop {} {\  (\bibinfo {year} {2006})},\ \Eprint
  {http://arxiv.org/abs/gr-qc/0604075} {arXiv:gr-qc/0604075 [gr-qc]}
  \BibitemShut {NoStop}%
\bibitem [{\citenamefont {Barcel\'o}\ \emph {et~al.}(2014)\citenamefont
  {Barcel\'o}, \citenamefont {Carballo-Rubio}, \citenamefont {Garay},\ and\
  \citenamefont {Jannes}}]{Barcelo:2014yna}%
  \BibitemOpen
  \bibfield  {author} {\bibinfo {author} {\bibfnamefont {C.}~\bibnamefont
  {Barcel\'o}}, \bibinfo {author} {\bibfnamefont {R.}~\bibnamefont
  {Carballo-Rubio}}, \bibinfo {author} {\bibfnamefont {L.~J.}\ \bibnamefont
  {Garay}}, \ and\ \bibinfo {author} {\bibfnamefont {G.}~\bibnamefont
  {Jannes}},\ }\href@noop {} {\  (\bibinfo {year} {2014})},\ \Eprint
  {http://arxiv.org/abs/1407.6532} {arXiv:1407.6532 [gr-qc]} \BibitemShut
  {NoStop}%
\bibitem [{\citenamefont {Unruh}(1981)}]{Unruh:1980cg}%
  \BibitemOpen
  \bibfield  {author} {\bibinfo {author} {\bibfnamefont {W.}~\bibnamefont
  {Unruh}},\ }\href {\doibase 10.1103/PhysRevLett.46.1351} {\bibfield
  {journal} {\bibinfo  {journal} {Phys.Rev.Lett.}\ }\textbf {\bibinfo {volume}
  {46}},\ \bibinfo {pages} {1351} (\bibinfo {year} {1981})}\BibitemShut
  {NoStop}%
\bibitem [{\citenamefont {Liberati}\ \emph {et~al.}(2009)\citenamefont
  {Liberati}, \citenamefont {Girelli},\ and\ \citenamefont
  {Sindoni}}]{Liberati:2009uq}%
  \BibitemOpen
  \bibfield  {author} {\bibinfo {author} {\bibfnamefont {S.}~\bibnamefont
  {Liberati}}, \bibinfo {author} {\bibfnamefont {F.}~\bibnamefont {Girelli}}, \
  and\ \bibinfo {author} {\bibfnamefont {L.}~\bibnamefont {Sindoni}},\
  }\href@noop {} {\  (\bibinfo {year} {2009})},\ \Eprint
  {http://arxiv.org/abs/0909.3834} {arXiv:0909.3834 [gr-qc]} \BibitemShut
  {NoStop}%
\bibitem [{\citenamefont {Barcelo}\ \emph {et~al.}(2005)\citenamefont
  {Barcelo}, \citenamefont {Liberati},\ and\ \citenamefont
  {Visser}}]{Barcelo:2005fc}%
  \BibitemOpen
  \bibfield  {author} {\bibinfo {author} {\bibfnamefont {C.}~\bibnamefont
  {Barcelo}}, \bibinfo {author} {\bibfnamefont {S.}~\bibnamefont {Liberati}}, \
  and\ \bibinfo {author} {\bibfnamefont {M.}~\bibnamefont {Visser}},\
  }\href@noop {} {\bibfield  {journal} {\bibinfo  {journal} {Living Rev.Rel.}\
  }\textbf {\bibinfo {volume} {8}},\ \bibinfo {pages} {12} (\bibinfo {year}
  {2005})},\ \Eprint {http://arxiv.org/abs/gr-qc/0505065} {arXiv:gr-qc/0505065
  [gr-qc]} \BibitemShut {NoStop}%
\bibitem [{\citenamefont {Jacobson}(1993)}]{Jacobson:1993hn}%
  \BibitemOpen
  \bibfield  {author} {\bibinfo {author} {\bibfnamefont {T.}~\bibnamefont
  {Jacobson}},\ }\href {\doibase 10.1103/PhysRevD.48.728} {\bibfield  {journal}
  {\bibinfo  {journal} {Phys.Rev.}\ }\textbf {\bibinfo {volume} {D48}},\
  \bibinfo {pages} {728} (\bibinfo {year} {1993})},\ \Eprint
  {http://arxiv.org/abs/hep-th/9303103} {arXiv:hep-th/9303103 [hep-th]}
  \BibitemShut {NoStop}%
\bibitem [{\citenamefont {Barcelo}\ \emph {et~al.}(2003)\citenamefont
  {Barcelo}, \citenamefont {Liberati},\ and\ \citenamefont
  {Visser}}]{Barcelo:2003wu}%
  \BibitemOpen
  \bibfield  {author} {\bibinfo {author} {\bibfnamefont {C.}~\bibnamefont
  {Barcelo}}, \bibinfo {author} {\bibfnamefont {S.}~\bibnamefont {Liberati}}, \
  and\ \bibinfo {author} {\bibfnamefont {M.}~\bibnamefont {Visser}},\ }\href
  {\doibase 10.1103/PhysRevA.68.053613} {\bibfield  {journal} {\bibinfo
  {journal} {Phys.Rev.}\ }\textbf {\bibinfo {volume} {A68}},\ \bibinfo {pages}
  {053613} (\bibinfo {year} {2003})},\ \Eprint
  {http://arxiv.org/abs/cond-mat/0307491} {arXiv:cond-mat/0307491 [cond-mat]}
  \BibitemShut {NoStop}%
\bibitem [{\citenamefont {Weinfurtner}\ \emph {et~al.}(2009)\citenamefont
  {Weinfurtner}, \citenamefont {Jain}, \citenamefont {Visser},\ and\
  \citenamefont {Gardiner}}]{Weinfurtner:2008if}%
  \BibitemOpen
  \bibfield  {author} {\bibinfo {author} {\bibfnamefont {S.}~\bibnamefont
  {Weinfurtner}}, \bibinfo {author} {\bibfnamefont {P.}~\bibnamefont {Jain}},
  \bibinfo {author} {\bibfnamefont {M.}~\bibnamefont {Visser}}, \ and\ \bibinfo
  {author} {\bibfnamefont {C.}~\bibnamefont {Gardiner}},\ }\href {\doibase
  10.1088/0264-9381/26/6/065012} {\bibfield  {journal} {\bibinfo  {journal}
  {Class.Quant.Grav.}\ }\textbf {\bibinfo {volume} {26}},\ \bibinfo {pages}
  {065012} (\bibinfo {year} {2009})},\ \Eprint {http://arxiv.org/abs/0801.2673}
  {arXiv:0801.2673 [gr-qc]} \BibitemShut {NoStop}%
\bibitem [{\citenamefont {Weinfurtner}\ \emph {et~al.}(2007)\citenamefont
  {Weinfurtner}, \citenamefont {Visser}, \citenamefont {Jain},\ and\
  \citenamefont {Gardiner}}]{Weinfurtner:2008ns}%
  \BibitemOpen
  \bibfield  {author} {\bibinfo {author} {\bibfnamefont {S.}~\bibnamefont
  {Weinfurtner}}, \bibinfo {author} {\bibfnamefont {M.}~\bibnamefont {Visser}},
  \bibinfo {author} {\bibfnamefont {P.}~\bibnamefont {Jain}}, \ and\ \bibinfo
  {author} {\bibfnamefont {C.}~\bibnamefont {Gardiner}},\ }\href@noop {}
  {\bibfield  {journal} {\bibinfo  {journal} {PoS}\ }\textbf {\bibinfo {volume}
  {QG-PH}},\ \bibinfo {pages} {044} (\bibinfo {year} {2007})},\ \Eprint
  {http://arxiv.org/abs/0804.1346} {arXiv:0804.1346 [gr-qc]} \BibitemShut
  {NoStop}%
\bibitem [{\citenamefont {Barcelo}\ \emph {et~al.}(2001)\citenamefont
  {Barcelo}, \citenamefont {Visser},\ and\ \citenamefont
  {Liberati}}]{Barcelo:2001tb}%
  \BibitemOpen
  \bibfield  {author} {\bibinfo {author} {\bibfnamefont {C.}~\bibnamefont
  {Barcelo}}, \bibinfo {author} {\bibfnamefont {M.}~\bibnamefont {Visser}}, \
  and\ \bibinfo {author} {\bibfnamefont {S.}~\bibnamefont {Liberati}},\ }\href
  {\doibase 10.1142/S0218271801001591} {\bibfield  {journal} {\bibinfo
  {journal} {Int.J.Mod.Phys.}\ }\textbf {\bibinfo {volume} {D10}},\ \bibinfo
  {pages} {799} (\bibinfo {year} {2001})},\ \Eprint
  {http://arxiv.org/abs/gr-qc/0106002} {arXiv:gr-qc/0106002 [gr-qc]}
  \BibitemShut {NoStop}%
\bibitem [{\citenamefont {Girelli}\ \emph {et~al.}(2009)\citenamefont
  {Girelli}, \citenamefont {Liberati},\ and\ \citenamefont
  {Sindoni}}]{Girelli:2008qp}%
  \BibitemOpen
  \bibfield  {author} {\bibinfo {author} {\bibfnamefont {F.}~\bibnamefont
  {Girelli}}, \bibinfo {author} {\bibfnamefont {S.}~\bibnamefont {Liberati}}, \
  and\ \bibinfo {author} {\bibfnamefont {L.}~\bibnamefont {Sindoni}},\ }\href
  {\doibase 10.1103/PhysRevD.79.044019} {\bibfield  {journal} {\bibinfo
  {journal} {Phys.Rev.}\ }\textbf {\bibinfo {volume} {D79}},\ \bibinfo {pages}
  {044019} (\bibinfo {year} {2009})},\ \Eprint {http://arxiv.org/abs/0806.4239}
  {arXiv:0806.4239 [gr-qc]} \BibitemShut {NoStop}%
\bibitem [{\citenamefont {Garay}\ \emph {et~al.}(2000)\citenamefont {Garay},
  \citenamefont {Anglin}, \citenamefont {Cirac},\ and\ \citenamefont
  {Zoller}}]{Garay:1999sk}%
  \BibitemOpen
  \bibfield  {author} {\bibinfo {author} {\bibfnamefont {L.}~\bibnamefont
  {Garay}}, \bibinfo {author} {\bibfnamefont {J.}~\bibnamefont {Anglin}},
  \bibinfo {author} {\bibfnamefont {J.}~\bibnamefont {Cirac}}, \ and\ \bibinfo
  {author} {\bibfnamefont {P.}~\bibnamefont {Zoller}},\ }\href {\doibase
  10.1103/PhysRevLett.85.4643} {\bibfield  {journal} {\bibinfo  {journal}
  {Phys.Rev.Lett.}\ }\textbf {\bibinfo {volume} {85}},\ \bibinfo {pages} {4643}
  (\bibinfo {year} {2000})},\ \Eprint {http://arxiv.org/abs/gr-qc/0002015}
  {arXiv:gr-qc/0002015 [gr-qc]} \BibitemShut {NoStop}%
\bibitem [{\citenamefont {Garay}\ \emph {et~al.}(2001)\citenamefont {Garay},
  \citenamefont {Anglin}, \citenamefont {Cirac},\ and\ \citenamefont
  {Zoller}}]{Garay:2000jj}%
  \BibitemOpen
  \bibfield  {author} {\bibinfo {author} {\bibfnamefont {L.}~\bibnamefont
  {Garay}}, \bibinfo {author} {\bibfnamefont {J.}~\bibnamefont {Anglin}},
  \bibinfo {author} {\bibfnamefont {J.}~\bibnamefont {Cirac}}, \ and\ \bibinfo
  {author} {\bibfnamefont {P.}~\bibnamefont {Zoller}},\ }\href {\doibase
  10.1103/PhysRevA.63.023611} {\bibfield  {journal} {\bibinfo  {journal}
  {Phys.Rev.}\ }\textbf {\bibinfo {volume} {A63}},\ \bibinfo {pages} {023611}
  (\bibinfo {year} {2001})},\ \Eprint {http://arxiv.org/abs/gr-qc/0005131}
  {arXiv:gr-qc/0005131 [gr-qc]} \BibitemShut {NoStop}%
\bibitem [{\citenamefont {Girelli}\ \emph {et~al.}(2008)\citenamefont
  {Girelli}, \citenamefont {Liberati},\ and\ \citenamefont
  {Sindoni}}]{Girelli:2008gc}%
  \BibitemOpen
  \bibfield  {author} {\bibinfo {author} {\bibfnamefont {F.}~\bibnamefont
  {Girelli}}, \bibinfo {author} {\bibfnamefont {S.}~\bibnamefont {Liberati}}, \
  and\ \bibinfo {author} {\bibfnamefont {L.}~\bibnamefont {Sindoni}},\ }\href
  {\doibase 10.1103/PhysRevD.78.084013} {\bibfield  {journal} {\bibinfo
  {journal} {Phys.Rev.}\ }\textbf {\bibinfo {volume} {D78}},\ \bibinfo {pages}
  {084013} (\bibinfo {year} {2008})},\ \Eprint {http://arxiv.org/abs/0807.4910}
  {arXiv:0807.4910 [gr-qc]} \BibitemShut {NoStop}%
\bibitem [{\citenamefont {Volovik}\ and\ \citenamefont
  {Zubkov}(2014)}]{Volovik:2013fca}%
  \BibitemOpen
  \bibfield  {author} {\bibinfo {author} {\bibfnamefont {G.}~\bibnamefont
  {Volovik}}\ and\ \bibinfo {author} {\bibfnamefont {M.}~\bibnamefont
  {Zubkov}},\ }\href {\doibase 10.1016/j.aop.2013.11.003} {\bibfield  {journal}
  {\bibinfo  {journal} {Annals Phys.}\ }\textbf {\bibinfo {volume} {340}},\
  \bibinfo {pages} {352} (\bibinfo {year} {2014})},\ \Eprint
  {http://arxiv.org/abs/1305.4665} {arXiv:1305.4665 [cond-mat.mes-hall]}
  \BibitemShut {NoStop}%
\bibitem [{\citenamefont {Volovik}(2009)}]{Volovik:2009av}%
  \BibitemOpen
  \bibfield  {author} {\bibinfo {author} {\bibfnamefont {G.}~\bibnamefont
  {Volovik}},\ }\href {\doibase 10.1134/S0021364009110010} {\bibfield
  {journal} {\bibinfo  {journal} {JETP Lett.}\ }\textbf {\bibinfo {volume}
  {89}},\ \bibinfo {pages} {525} (\bibinfo {year} {2009})},\ \Eprint
  {http://arxiv.org/abs/0904.4113} {arXiv:0904.4113 [gr-qc]} \BibitemShut
  {NoStop}%
\bibitem [{\citenamefont {Jannes}\ and\ \citenamefont
  {Volovik}(2012)}]{Jannes:2011em}%
  \BibitemOpen
  \bibfield  {author} {\bibinfo {author} {\bibfnamefont {G.}~\bibnamefont
  {Jannes}}\ and\ \bibinfo {author} {\bibfnamefont {G.}~\bibnamefont
  {Volovik}},\ }\href {\doibase 10.1134/S0021364012160035} {\bibfield
  {journal} {\bibinfo  {journal} {JETP Lett.}\ }\textbf {\bibinfo {volume}
  {96}},\ \bibinfo {pages} {215} (\bibinfo {year} {2012})},\ \Eprint
  {http://arxiv.org/abs/1108.5086} {arXiv:1108.5086 [gr-qc]} \BibitemShut
  {NoStop}%
\bibitem [{\citenamefont {Sindoni}\ \emph {et~al.}(2009)\citenamefont
  {Sindoni}, \citenamefont {Girelli},\ and\ \citenamefont
  {Liberati}}]{Sindoni:2009fc}%
  \BibitemOpen
  \bibfield  {author} {\bibinfo {author} {\bibfnamefont {L.}~\bibnamefont
  {Sindoni}}, \bibinfo {author} {\bibfnamefont {F.}~\bibnamefont {Girelli}}, \
  and\ \bibinfo {author} {\bibfnamefont {S.}~\bibnamefont {Liberati}},\
  }\href@noop {} {\  (\bibinfo {year} {2009})},\ \Eprint
  {http://arxiv.org/abs/0909.5391} {arXiv:0909.5391 [gr-qc]} \BibitemShut
  {NoStop}%
\bibitem [{\citenamefont {Finazzi}\ \emph {et~al.}(2012)\citenamefont
  {Finazzi}, \citenamefont {Liberati},\ and\ \citenamefont
  {Sindoni}}]{Finazzi:2011zw}%
  \BibitemOpen
  \bibfield  {author} {\bibinfo {author} {\bibfnamefont {S.}~\bibnamefont
  {Finazzi}}, \bibinfo {author} {\bibfnamefont {S.}~\bibnamefont {Liberati}}, \
  and\ \bibinfo {author} {\bibfnamefont {L.}~\bibnamefont {Sindoni}},\ }\href
  {\doibase 10.1103/PhysRevLett.108.071101} {\bibfield  {journal} {\bibinfo
  {journal} {Phys.Rev.Lett.}\ }\textbf {\bibinfo {volume} {108}},\ \bibinfo
  {pages} {071101} (\bibinfo {year} {2012})},\ \Eprint
  {http://arxiv.org/abs/1103.4841} {arXiv:1103.4841 [gr-qc]} \BibitemShut
  {NoStop}%
\bibitem [{\citenamefont {Fagnocchi}\ \emph {et~al.}(2010)\citenamefont
  {Fagnocchi}, \citenamefont {Finazzi}, \citenamefont {Liberati}, \citenamefont
  {Kormos},\ and\ \citenamefont {Trombettoni}}]{Fagnocchi:2010sn}%
  \BibitemOpen
  \bibfield  {author} {\bibinfo {author} {\bibfnamefont {S.}~\bibnamefont
  {Fagnocchi}}, \bibinfo {author} {\bibfnamefont {S.}~\bibnamefont {Finazzi}},
  \bibinfo {author} {\bibfnamefont {S.}~\bibnamefont {Liberati}}, \bibinfo
  {author} {\bibfnamefont {M.}~\bibnamefont {Kormos}}, \ and\ \bibinfo {author}
  {\bibfnamefont {A.}~\bibnamefont {Trombettoni}},\ }\href {\doibase
  10.1088/1367-2630/12/9/095012} {\bibfield  {journal} {\bibinfo  {journal}
  {New J.Phys.}\ }\textbf {\bibinfo {volume} {12}},\ \bibinfo {pages} {095012}
  (\bibinfo {year} {2010})},\ \Eprint {http://arxiv.org/abs/1001.1044}
  {arXiv:1001.1044 [gr-qc]} \BibitemShut {NoStop}%
\bibitem [{\citenamefont {Wald}(1984)}]{Wald:1984rg}%
  \BibitemOpen
  \bibfield  {author} {\bibinfo {author} {\bibfnamefont {R.~M.}\ \bibnamefont
  {Wald}},\ }\href@noop {} {\emph {\bibinfo {title} {{General Relativity}}}}\
  (\bibinfo {year} {1984})\BibitemShut {NoStop}%
\bibitem [{\citenamefont {Kapusta}(1981)}]{Kapusta:1981aa}%
  \BibitemOpen
  \bibfield  {author} {\bibinfo {author} {\bibfnamefont {J.~I.}\ \bibnamefont
  {Kapusta}},\ }\href {\doibase 10.1103/PhysRevD.24.426} {\bibfield  {journal}
  {\bibinfo  {journal} {Phys.Rev.}\ }\textbf {\bibinfo {volume} {D24}},\
  \bibinfo {pages} {426} (\bibinfo {year} {1981})}\BibitemShut {NoStop}%
\bibitem [{\citenamefont {Bernstein}\ and\ \citenamefont
  {Dodelson}(1991)}]{Bernstein:1990kf}%
  \BibitemOpen
  \bibfield  {author} {\bibinfo {author} {\bibfnamefont {J.}~\bibnamefont
  {Bernstein}}\ and\ \bibinfo {author} {\bibfnamefont {S.}~\bibnamefont
  {Dodelson}},\ }\href {\doibase 10.1103/PhysRevLett.66.683} {\bibfield
  {journal} {\bibinfo  {journal} {Phys.Rev.Lett.}\ }\textbf {\bibinfo {volume}
  {66}},\ \bibinfo {pages} {683} (\bibinfo {year} {1991})}\BibitemShut
  {NoStop}%
\bibitem [{\citenamefont {Haber}\ and\ \citenamefont
  {Weldon}(1982)}]{Haber:1981ts}%
  \BibitemOpen
  \bibfield  {author} {\bibinfo {author} {\bibfnamefont {H.~E.}\ \bibnamefont
  {Haber}}\ and\ \bibinfo {author} {\bibfnamefont {H.~A.}\ \bibnamefont
  {Weldon}},\ }\href {\doibase 10.1103/PhysRevD.25.502} {\bibfield  {journal}
  {\bibinfo  {journal} {Phys.Rev.}\ }\textbf {\bibinfo {volume} {D25}},\
  \bibinfo {pages} {502} (\bibinfo {year} {1982})}\BibitemShut {NoStop}%
\bibitem [{\citenamefont {Haber}\ and\ \citenamefont
  {Weldon}(1981)}]{Haber:1981fg}%
  \BibitemOpen
  \bibfield  {author} {\bibinfo {author} {\bibfnamefont {H.~E.}\ \bibnamefont
  {Haber}}\ and\ \bibinfo {author} {\bibfnamefont {H.~A.}\ \bibnamefont
  {Weldon}},\ }\href {\doibase 10.1103/PhysRevLett.46.1497} {\bibfield
  {journal} {\bibinfo  {journal} {Phys.Rev.Lett.}\ }\textbf {\bibinfo {volume}
  {46}},\ \bibinfo {pages} {1497} (\bibinfo {year} {1981})}\BibitemShut
  {NoStop}%
\bibitem [{\citenamefont {Pitaevskii}\ and\ \citenamefont
  {Stringari}(2003)}]{pitaevskii2003bose}%
  \BibitemOpen
  \bibfield  {author} {\bibinfo {author} {\bibfnamefont {L.}~\bibnamefont
  {Pitaevskii}}\ and\ \bibinfo {author} {\bibfnamefont {S.}~\bibnamefont
  {Stringari}},\ }\href@noop {} {\emph {\bibinfo {title} {Bose-Einstein
  Condensation}}}\ (\bibinfo {year} {2003})\BibitemShut {NoStop}%
\bibitem [{\citenamefont {Visser}\ and\ \citenamefont
  {Molina-Paris}(2010)}]{Visser:2010xv}%
  \BibitemOpen
  \bibfield  {author} {\bibinfo {author} {\bibfnamefont {M.}~\bibnamefont
  {Visser}}\ and\ \bibinfo {author} {\bibfnamefont {C.}~\bibnamefont
  {Molina-Paris}},\ }\href {\doibase 10.1088/1367-2630/12/9/095014} {\bibfield
  {journal} {\bibinfo  {journal} {New J.Phys.}\ }\textbf {\bibinfo {volume}
  {12}},\ \bibinfo {pages} {095014} (\bibinfo {year} {2010})},\ \Eprint
  {http://arxiv.org/abs/1001.1310} {arXiv:1001.1310 [gr-qc]} \BibitemShut
  {NoStop}%
\bibitem [{\citenamefont {Bilic}(1999)}]{Bilic:1999sq}%
  \BibitemOpen
  \bibfield  {author} {\bibinfo {author} {\bibfnamefont {N.}~\bibnamefont
  {Bilic}},\ }\href {\doibase 10.1088/0264-9381/16/12/312} {\bibfield
  {journal} {\bibinfo  {journal} {Class.Quant.Grav.}\ }\textbf {\bibinfo
  {volume} {16}},\ \bibinfo {pages} {3953} (\bibinfo {year} {1999})},\ \Eprint
  {http://arxiv.org/abs/gr-qc/9908002} {arXiv:gr-qc/9908002 [gr-qc]}
  \BibitemShut {NoStop}%
\bibitem [{\citenamefont {Moncrief}(1980)}]{Moncrief}%
  \BibitemOpen
  \bibfield  {author} {\bibinfo {author} {\bibfnamefont {V.}~\bibnamefont
  {Moncrief}},\ }\href@noop {} {\bibfield  {journal} {\bibinfo  {journal}
  {Astrophys. J.}\ }\textbf {\bibinfo {volume} {235}},\ \bibinfo {pages} {1038}
  (\bibinfo {year} {1980})}\BibitemShut {NoStop}%
\bibitem [{\citenamefont {Bilic}\ and\ \citenamefont
  {Tolic}(2013)}]{Bilic:2013qpa}%
  \BibitemOpen
  \bibfield  {author} {\bibinfo {author} {\bibfnamefont {N.}~\bibnamefont
  {Bilic}}\ and\ \bibinfo {author} {\bibfnamefont {D.}~\bibnamefont {Tolic}},\
  }\href {\doibase 10.1103/PhysRevD.88.105002} {\bibfield  {journal} {\bibinfo
  {journal} {Phys.Rev.}\ }\textbf {\bibinfo {volume} {D88}},\ \bibinfo {pages}
  {105002} (\bibinfo {year} {2013})},\ \Eprint {http://arxiv.org/abs/1309.2833}
  {arXiv:1309.2833 [gr-qc]} \BibitemShut {NoStop}%
\bibitem [{\citenamefont {Deruelle}(2011)}]{Deruelle:2011wu}%
  \BibitemOpen
  \bibfield  {author} {\bibinfo {author} {\bibfnamefont {N.}~\bibnamefont
  {Deruelle}},\ }\href {\doibase 10.1007/s10714-011-1247-x} {\bibfield
  {journal} {\bibinfo  {journal} {Gen.Rel.Grav.}\ }\textbf {\bibinfo {volume}
  {43}},\ \bibinfo {pages} {3337} (\bibinfo {year} {2011})},\ \Eprint
  {http://arxiv.org/abs/1104.4608} {arXiv:1104.4608 [gr-qc]} \BibitemShut
  {NoStop}%
\bibitem [{\citenamefont {Giulini}(2006)}]{Giulini:2006ry}%
  \BibitemOpen
  \bibfield  {author} {\bibinfo {author} {\bibfnamefont {D.}~\bibnamefont
  {Giulini}},\ }\href@noop {} {\  (\bibinfo {year} {2006})},\ \Eprint
  {http://arxiv.org/abs/gr-qc/0611100} {arXiv:gr-qc/0611100 [gr-qc]}
  \BibitemShut {NoStop}%
\bibitem [{\citenamefont {Di~Casola}\ \emph {et~al.}(2014)\citenamefont
  {Di~Casola}, \citenamefont {Liberati},\ and\ \citenamefont
  {Sonego}}]{DiCasola:2013yga}%
  \BibitemOpen
  \bibfield  {author} {\bibinfo {author} {\bibfnamefont {E.}~\bibnamefont
  {Di~Casola}}, \bibinfo {author} {\bibfnamefont {S.}~\bibnamefont {Liberati}},
  \ and\ \bibinfo {author} {\bibfnamefont {S.}~\bibnamefont {Sonego}},\ }\href
  {\doibase 10.1103/PhysRevD.89.084053} {\bibfield  {journal} {\bibinfo
  {journal} {Phys.Rev.}\ }\textbf {\bibinfo {volume} {D89}},\ \bibinfo {pages}
  {084053} (\bibinfo {year} {2014})},\ \Eprint {http://arxiv.org/abs/1401.0030}
  {arXiv:1401.0030 [gr-qc]} \BibitemShut {NoStop}%
\bibitem [{\citenamefont {Weinberg}\ and\ \citenamefont
  {Witten}(1980)}]{20623}%
  \BibitemOpen
  \bibfield  {author} {\bibinfo {author} {\bibfnamefont {S.}~\bibnamefont
  {Weinberg}}\ and\ \bibinfo {author} {\bibfnamefont {E.}~\bibnamefont
  {Witten}},\ }\href {\doibase 10.1016/0370-2693(80)90212-9} {\bibfield
  {journal} {\bibinfo  {journal} {Phys. Lett. B}\ }\textbf {\bibinfo {volume}
  {96}},\ \bibinfo {pages} {59} (\bibinfo {year} {1980})}\BibitemShut {NoStop}%
\bibitem [{\citenamefont {Oriti}(2014)}]{Oriti:2013jga}%
  \BibitemOpen
  \bibfield  {author} {\bibinfo {author} {\bibfnamefont {D.}~\bibnamefont
  {Oriti}},\ }\href {\doibase 10.1016/j.shpsb.2013.10.006} {\bibfield
  {journal} {\bibinfo  {journal} {Stud.Hist.Philos.Mod.Phys.}\ }\textbf
  {\bibinfo {volume} {46}},\ \bibinfo {pages} {186} (\bibinfo {year} {2014})},\
  \Eprint {http://arxiv.org/abs/1302.2849} {arXiv:1302.2849 [physics.hist-ph]}
  \BibitemShut {NoStop}%
\end{thebibliography}%
\end{document}